\documentclass[sigconf]{acmart}

\usepackage{multirow}
\usepackage{graphicx}
\usepackage{soul}
\usepackage[normalem]{ulem}
\usepackage{url}
\AtBeginDocument{%
  }

\copyrightyear{2025}
\acmYear{2025}
\setcopyright{rightsretained}
\acmConference[CSCW Companion '25]{Companion of the Computer-Supported Cooperative Work and Social Computing}{October 18--22, 2025}{Bergen, Norway}
\acmBooktitle{Companion of the Computer-Supported Cooperative Work and Social Computing (CSCW Companion '25), October 18--22, 2025, Bergen, Norway}\acmDOI{10.1145/3715070.3749230}
\acmISBN{979-8-4007-1480-1/2025/10}

\begin{document}

\title[From Platform Migration to Cultural Integration: the Ingress and Diffusion of \#wlw \\from TikTok to RedNote in Queer Women Communities]{From Platform Migration to Cultural Integration: the Ingress and Diffusion of \#wlw from TikTok to RedNote in Queer Women Communities}

\author{Ziqi Pan}
\email{zpanar@connect.ust.hk}
\orcid{0000-0002-5562-8685}
\affiliation{%
  \institution{The Hong Kong University of Science and Technology}
  \city{Hong Kong}
  \country{China}
}

\author{Runhua Zhang}
\email{runhua.zhang@connect.ust.hk}
\orcid{0000-0002-0519-5148}
\affiliation{%
  \institution{The Hong Kong University of Science and Technology}
  \city{Hong Kong}
  \country{China}
}

\author{Jiehui Luo}
\email{jluo@nd.edu}
\orcid{0000-0002-1394-4348}
\affiliation{%
  \institution{University of Notre Dame}
  \city{South Bend}
  \country{United States}
}

\author{Yuanhao Zhang}
\email{yzhangiy@connect.ust.hk}
\orcid{0000-0001-8263-1823}
\affiliation{%
  \institution{The Hong Kong University of Science and Technology}
  \city{Hong Kong}
  \country{China}
}

\author{Yue Deng}
\email{ydengbi@connect.ust.hk}
\orcid{0009-0008-5339-8792}
\affiliation{%
  \institution{The Hong Kong University of Science and Technology}
  \city{Hong Kong}
  \country{China}
}

\author{Xiaojuan MA}
\email{mxj@cse.ust.hk}
\orcid{0000-0002-9847-7784}
\affiliation{%
  \institution{The Hong Kong University of Science and Technology}
  \city{Hong Kong}
  \country{China}
}

\renewcommand{\shortauthors}{Pan et al.}

\begin{abstract}
Hashtags serve as identity markers and connection tools in online queer communities. Recently, the Western-origin \#wlw (women-loving-women) hashtag has risen in the Chinese lesbian community on RedNote, coinciding with user migration triggered by the temporary US TikTok ban. This event provides a unique lens to study cross-cultural hashtag ingress and diffusion through the populations' responsive behaviors in cyber-migration. In this paper, we conducted a two-phase content analysis of 418 \#wlw posts from January and April, examining different usage patterns during the hashtag's ingress and diffusion. Results indicate that the successful introduction of \#wlw was facilitated by TikTok immigrants' bold importation, both populations' mutual interpretation, and RedNote natives' discussions. In current manifestation of diffusion, \#wlw becomes a RedNote-recognized queer hashtag for sharing queer life, and semantically expands to support feminism discourse. Our findings provide empirical insights for enhancing the marginalized communities' cross-cultural communication.
\end{abstract}




\maketitle

\section{Introduction and Background}
Queer people are longing for shelters in the cyberspace \cite{wagner_social_2024}, where expressive hashtags, are frequently employed by queer users on all-audience social media to represent their identities and connect with relevant communities \cite{ananthasubramaniam_roles_2025}. 
%Online queer communities witness various hashtags' birth through lifecycle even to death. 
In earlier practices, explicit sexuality hashtags such as \#lesbian and \#IAmGay were popular for queer identity disclosure \cite{herrera_theorizing_2018, liao2019iamgay}, but their limited inclusivity sparked controversy and eventually decreased their usage \cite{herrera_theorizing_2018}. As a result, more implicit and inclusive tags like \#loveislove and rainbow emojis became favored in online queer spaces \cite{zhang_hashtagging_2024, la2020lgbtqi+}. Among these evolving forms, \#wlw (women-loving-women) emerged in the late 2010s as a relatively safe and inclusive hashtag that embraces various identities under the queer female spectrum, including lesbian, bisexual, pansexual, and transgender individuals \cite{wlw_dating, herrera_theorizing_2018}. Gradually, it becomes a widely used and emotionally resonant label across Western social media platforms like TikTok \cite{tiktok_new}. 

Recently, the hashtag \#wlw entered the Chinese lesbian community on RedNote, following a wave of TikTok users seeking alternatives during the U.S. TikTok ban \cite{yuan2025love}.
Prior to this migration, \#wlw had only ``several thousand views'' \footnote{\url{https://shorturl.at/bzS4J}} on RedNote. However, within just two days of the immigrant influx, \#wlw soared to 47 million views and remained active even after the majority of TikTok users returned following the ban's lift \cite{enwiki:1289171792}. As of May 12, 2025, the hashtag had reached 360 million views \footnote{\url{https://www.xiaohongshu.com/topic/normal/616ec6da6292a700017c9c69}}.
This dramatic rise and sustained popularity of \#wlw presents a unique case of hashtag ingress, diffusion, and localization in a cross-cultural context. It offers a valuable opportunity to examine how queer-coded digital language evolves across platforms and cultures, which would offer 
insights into understanding effective communication across diverse communities, specifically among queer people.

However, limited studies have examined such cross-cultural hashtag ingress and diffusion as a reactive process driven by a certain event, like cyber-migration.
While prior HCI and CSCW research has extensively explored hashtag development at both macro and micro levels, the focus has largely been on proactive hashtag use within single platforms. 
At the macro level, studies have studied the overall pattern of hashtag developments, such as patterns of information spread \cite{6921598}, and temporal and spatial dynamics of user-created subgraphs around hashtags \cite{ardon2013spatio}. 
Micro-level research has examined how users and platforms create and utilize specific hashtags. 
Studies have shown that users employ hashtags for activism, supporting movements such as \#MeToo and \#BlackLivesMatter \cite{wang2023e, rho2020, rho2019, hansson2021organizing}, as well as for increasing the visibility of marginalized groups like disabled communities \cite{sannon2023disability, kaur2024challenges}. 
Researchers have also explored how hashtags contribute to building sub-communities based on users’ positionalities, such as ethnicity \cite{lutz2024we}, gender \cite{wan2025a, hasebrink2018analysing}, sexuality \cite{llewellyn2022space}, and disability \cite{sannon2023disability, kaur2024challenges, mcdonnell2024caption}.
Notably, Wan et al. \cite{wan2025a} investigated hashtag re-appropriation, where users construct sub-communities by obscuring content visibility from unwanted audiences rather than attracting desired ones.
Additionally, research discusses how platforms support sub-community operation through hashtag-based content curation \cite{lutz2024we, kender2025social, souza2025mediating}. Within queer communities specifically, hashtags have been studied as important tools for identity representation \cite{freeman2020streaming, oakley2016disturbing}, building connections \cite{carrasco2018queer}, and fostering inclusion \cite{zhang_hashtagging_2024}. 
Yet, few studies have focused on hashtags' role in cross-cultural communication of queer communities.
This gap is especially notable as cross-cultural queer communication is crucial for supporting peers and amplifying marginalized voices \cite{chavez2013pushing, eguchi2021horizon, cui2022we}, yet it is often hindered by different cultural norms and digital practices across queer communities \cite{zottola2024lgbtq+, berger2022social}.
Understanding how hashtags migrate and evolve across cultural boundaries-especially in queer online communities-can inform strategies for strengthening cross-cultural connection, promoting digital equity, and amplifying underrepresented voices.

In this paper, we address this gap by examining the ingress and diffusion of the \#wlw hashtag on RedNote after its migration from TikTok.
Specifically, we investigated this through two research questions: \textbf{RQ1: How did the TikTok immigrants and Rednote natives use the \#wlw upon the hashtag ingress when influx occurred?} and \textbf{RQ2: What is the current manifestation of hashtag diffusion indicated by the usage of \#wlw on Rednote after the efflux of immigrants?} 
To address the RQs, we collected a dataset of 418 popular posts featuring the \#wlw hashtag on RedNote, sampled at two temporal points: January and April 2025. We then conducted a content analysis on the posts from both temporal points, with each corresponding to one of the research questions.

To address RQ1, we analyzed the content composition of 201 posts from January, capturing the usage patterns during the influx period. Our analysis reveals that the successful introduction of \#wlw was a collaborative effort between immigrants and natives. 
As immigrants transplanted TikTok’s \#wlw usage patterns, such as self-promotion for socializing, to RedNote, both groups adopted \#wlw alongside native lesbian hashtags to interpret it functionally. Meanwhile, natives discussed the literal meaning of \#wlw, signaling a negotiation of its relevance in the local context.

To address RQ2, we examined 217 posts from April and compared their content composition with the January dataset to identify the shifts in usage over time. We observed a decline in socializing content (\textit{e.g.}, dating requests and friend-making), alongside a rise in diverse sharing practices (\textit{e.g.}, lesbian couple life, lesbian humor, and instructions for lesbians).
This suggests that \#wlw, as a newly recognized lesbian hashtag, has adapted to RedNote's role as a platform for sharing life in the Chinese online lesbian ecosystem after the influx-induced friend-making trend ceases. 
Notably, the rise in gender and feminist discussions using \#wlw in April marks its semantic expansion from a Western queer identifier to a locally integrated term within native lesbian and feminist discourse.

%conclusion
Our paper's contributions are as follows: First, our findings reveal how sub-communities from different culture backgrounds communicate in cyberspace through hashtags as multifaceted mediators, with the example of \#wlw in facilitating the interaction and adaptation between Western and Chinese queer users. Second, we offer valuable insights on how social media platforms can better support cross-cultural communication, especially for marginalized groups.

\section{Method}

\subsection{Data Collection}
Upon noticing a surge in \#wlw posts on RedNote alongside TikTok immigrants' influx on January 15, we started a virtual ethnographic study \cite{hine_virtual_2025}. Through the study, we identified distinct patterns in \#wlw usage between its initial introduction and later diffusion, which guided us to select two temporal data collection points to further study the usage patterns.
Using a hashtag-based search with \#wlw and newly created accounts to eliminate recommendation bias, we retrieved the most popular posts sorted by RedNote's default parameters via MediaCrawler \footnote{\url{https://github.com/NanmiCoder/MediaCrawler?tab=readme-ov-file}}.
We avoided time-based sorting when collecting due to RedNote's restriction on collecting continuous temporal data, which could introduce bias.
Data collection occurred in two phases, January and April 2025, corresponding to the ingress and post-diffusion stages. Initially, we collected 460 posts constrained to RedNote's search limit. After removing duplicates and unavailable posts, we obtained a final dataset comprising 201 posts from January and 217 from April. For each post, we gathered detailed content data, including titles, text descriptions, images/videos, and metadata such as user ID, IP location, hashtags, URLs, timestamps, and engagement metrics (comments, views, and upvotes). 
While not exhaustive, the dataset allows us to identify recurring themes and generate preliminary insights for future research on cross-cultural digital practices.

\subsection{Data Analysis}
To address our research questions, we started with data preprocessing. We labeled authors as TikTok immigrants or RedNote natives based on IP locations and posting history, categorized posts based on hashtag composition by the presence of RedNote native tags (\textit{e.g.}, \#le) and the foreign \#wlw tag. 
Then, we conducted a content analysis \cite{drisko2016content} of \#wlw posts during the cyber-immigrant influx and after the efflux periods. 
Adopting an inductive approach \cite{hadi2022gamification, hadi2022users}, four authors independently coded 40 randomly selected posts, then collectively developed an initial codebook.
This codebook was validated by independently analyzing 60 additional posts, with discrepancies resolved through discussion.
The resulting 17-code codebook guided analysis of the remaining 318 posts, with flexibility to incorporate emergent patterns. Two additional codes were added, finalizing a 19-code codebook.
Each post was reviewed by at least two coders, with disagreements resolved through consensus.
Finally, these codes were synthesized into higher-level themes.
For RQ1, we analyzed the content composition of January posts by examining the distribution of themes and codes.
For RQ2, we performed a similar analysis on April posts and compared their content composition to that of January.

\section{Findings}
% 110/201 216/217
Our analysis yielded six themes regarding the content of the posts: Commercial Content, Influx Related Dialogues, Knowledgeable Content, Personal Showcase Content, Recreational Content, and Socializing Content, each containing subordinate codes and subcodes (see Figure \ref{fig:distribution} for themes and codes; complete codebook in Appendix Table \ref{tab:coding}). The distribution of posts by theme and code for each stage is visualized in Figure \ref{fig:distribution}.
% code book

\begin{figure*}[h]
  \includegraphics[width=0.7\textwidth]{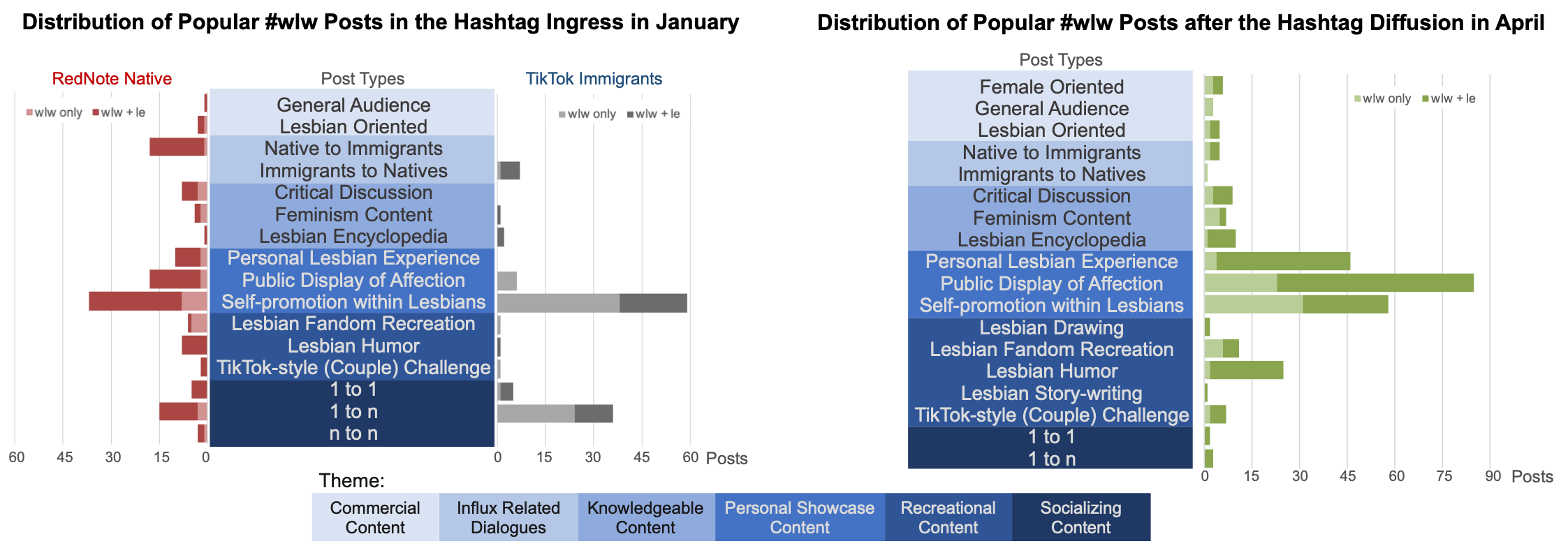}
  \caption{Distribution of Popular \#wlw Posts Based on Their Content Type in January and April}
  \Description{Distribution of Popular \#wlw Posts Based on Their Content Type in January and April}
  \label{fig:distribution}
\end{figure*}

\subsection{Upon Ingress: Bold Landing, Mutual Interpretation, and Passionate Learning (RQ1)}

\subsubsection{Natural Importation and Transplantation of \#wlw Hashtag Usage From TikTok to RedNote by Immigrants}
TikTok immigrants predominantly posted content (91 posts in total) of two main themes: \textbf{\textit{Personal Showcase Content}} (67/91) and \textbf{\textit{Socializing Content}} (44/91), reflecting their desire to establish connections within their sub-community on RedNote.
Of the 67 \textbf{\textit{Personal Showcase Content}} posts, 59 featured TikTok-style self-promotions (including look checks, lip-syncing while displaying appearance, and thirst traps). Significantly, 23 of these posts incorporated explicit socializing elements, with users either directly stating intentions like ``\textit{sincerely finding friends}" in descriptions or employing networking hashtags such as \#moots (shorthand for "mutuals") to attract followers.
This phenomenon illustrates how immigrants naturally carried over their posting habits from TikTok to establish the practices of posting self-promotional content with \#wlw on RedNote and effectively used this content to network with new audiences.

\subsubsection{Mutual Interpretation of \#wlw by Immigrants and Natives Leveraging RedNote Native Hashtags}\label{misuse}
After immigrants introduced \#wlw, both groups contributed to its cross-cultural interpretation and usage, leveraging RedNote-style lesbian hashtags (\textit{e.g.}, \#le, \#Le) as a bridge in this process.
While only three posts explicitly questioned the meaning of \#wlw and its relation to \#le, a striking 41 out of 91 immigrant posts and 79 out of 110 native posts used both hashtags, demonstrating that the populations found the two hashtags to be synonyms in terms of practical usage before literal meaning.
While immigrants introduced \#wlw to RedNote with the explanatory \#le, natives more broadly combined and popularized both.
It is worth noting that despite their functional convergence in usage, the terms maintain distinct ontological meanings within queer discourse: \#wlw represents an inclusive umbrella term encompassing lesbian, bisexual, and pansexual women, while \#le specifically denotes lesbian identity.

\subsubsection{Understanding \#wlw by Its Literal Meaning through Discussions among RedNote Natives}
% interpretation by synonym can guide the usage of wlw, others look into more ontological definition of wlw
% Some expressed affective reflection of wlw
% Some WLW promoters start serious discussions
The interpretation of \#wlw through the native \#le provided practical usage guidance, even without complete conceptual understanding. However, many natives still aim to explore the conceptual meaning of \#wlw within the queer knowledge framework. 
Some expressed their feelings about \#wlw, noting how it helped them feel included in the women-loving community on RedNote. 
A pansexual user said, \textit{``As pansexual, I feel even more visible than lesbians... \#wlw made me feel welcome''}. Another commented, \textit{``Bisexuals have a way out.''}
Some users further discussed such inclusion. A \#wlw activist remarked, \textit{``What we couldn't do for inclusiveness, perhaps Western queers can—since their slang is now on Chinese social media''}. Another noted, \textit{``Lesbian tags in China are more stereotyped, but love shouldn’t be confined by labels.''}

\subsection{Current Manifestation of the Hashtag Diffusion: Fostering \#wlw with Local Culture (RQ2)}
\subsubsection{Overview of the Content Composition}
In April, we found that the majority of creators (216 out of 217) were natives, while the daily views and discussion activity for \#wlw remained around 5 million and 30,000 (derived from the \#wlw homepage\footnote{\url{https://www.xiaohongshu.com/topic/normal/616ec6da6292a700017c9c69}}). 
These statistics suggest that, although TikTok immigrants have faded from RedNote, their \#wlw hashtag resides on RedNote and becomes widely recognized as a lesbian marker within the native community.
\textbf{\textit{Personal Showcase Content}} still dominated \#wlw usage, with the top three post types being public displays of affection, self-promotion, and personal experience sharing. Following \textbf{\textit{Personal Showcase Content}}, \textbf{\textit{Recreational Content}}, and \textbf{\textit{Knowledgeable Content}} became the next two themes that showed popularity in the current usage of \#wlw.
% First based on IP, immigrants left.
% Personal showcase content dominating，the composition of showcase content: Public affection > Self-promotion > personal experience
% Recreational content and knowledgeable content followed

\subsubsection{Shifting back from Friend-making Market to Content-sharing Radio}
% mainly talk about: (1) drastic decrease in friend-making content (2) the shift in personal showcase content from single to couple
% 功能从与ppl with same sexual orientation 交友 稳定为 向指定目标人群分享
% 暗合了小红书作为一个“分享”性质平台的nature
% 一定程度上印证了wlw成为了一个被local社群也认证的能够标记lesbian群体的tag
Compared to January, \textbf{\textit{Socializing Content}} dropped sharply from 64/201 to 5/217 posts. While \textbf{\textit{Personal Showcase}} remained the dominant theme, its focus shifted: self-promotion, which led in January (96/130), was overtaken in April by public displays of lesbian couples' affection (from 24/130 to 85/189), followed by self-promotion (58/189) and experience sharing (46/189). \textbf{Recreational Content} also grew, diversifying \#wlw through humor, fandom, and art contents.
This shift highlights that as the immigrant-driven friend-seeking wave waned, native RedNote users redefined posting featuring \#wlw as a way to share daily life with targeted audiences, aligning with RedNote’s native characteristics and role of sharing rather than dating or networking. 

\subsubsection{Extended Usage of \#wlw: From Lesbian Instructions to Beyond Queer Topics}
The increase in \textbf{\textit{Knowledgeable Content}} is remarkable, rising from 16 out of 201 posts to 26 out of 217. Of these, ten were categorized as lesbian encyclopedia entries, with eight specifically classified as instructional content for lesbians—a subset emerging exclusively in April's dataset.
One influencer posted a video series guiding new lesbians on sexuality and dating, while others shared relationship advice from personal experience.
\#Wlw has also extended beyond queer discourse into broader feminist conversations, appearing in seven feminism-related posts.
Notably, the platform's most popular \#wlw-tagged feminist post—a satirical critique of workplace gender inequality—garnered 275k likes, suggesting both RedNote's receptiveness to feminist content and potential community overlap between feminist and lesbian audiences.
Interestingly, the focus of feminist topics featuring \#wlw varies depending on accompanying hashtags. Posts with only \#wlw tend to discuss general gender issues in daily life, while those combining \#le and \#wlw concentrate on gender dynamics specific to lesbian experiences, such as heterosexual norms affecting the division of roles within lesbian couples \cite{huxley2014resisting, ochse2011real}. 

\section{Discussion and Future Works}
Initially, \#wlw rapidly ingressed RedNote through TikTok immigrants' proactive transplantation, and was embraced by natives.
This response from natives reflected both positive reception and collective curiosity toward the hashtag as a more expressive and celebratory symbol of queer identity.
The sudden visibility of Western lesbians stimulated the dormant vitality of Chinese lesbian users, who had long hidden behind implicit hashtags \cite{shen2023safe, cui2022we, wen2025exploring}. For example, they use a common English suffix \#le to abbreviate ``lesbian'', and a common Chinese name \#chenle to refer to tomboy lesbians. 
Western immigrants, unfamiliar with these coded norms, introduced a more open and explicit style of queer expression, unintentionally provoking and inspiring native users to adopt \#wlw and express themselves more directly in the women-loving-women context. 
This phenomenon shows that \textbf{cross-cultural queer communication goes beyond dissemination of language or symbols; it also reshape the identity politics and self-expression approaches in other communities}. 
During the diffusion of \#wlw on RedNote, although native lesbian tags aided functional interpretation, confusion over the literal meanings of hashtags arose (Section \ref{misuse}).
Such misuse of hashtag synonyms may lead to communicative misalignment and diminish the potential for mutual cultural enrichment between the queer communities. This calls for platforms to \textbf{enhance cross-cultural understanding by minimizing misalignment of shared digital language}, such as via integrating translation tools or explanatory plugins.

As the TikTok immigrants effluxed, \#wlw diffused and underwent semantic broadening within the RedNote culture.
Originally a Western queer label, \#wlw expanded its semantic scope to integrate local lesbian terminology and feminist discourse.
On one hand, \#wlw adapted to RedNote's platform culture and specialty of ``recommending and sharing life,'' \cite{enwiki:1288274166} evolving into a playground for sharing more diverse queer content (\textit{e.g.}, lesbian relationship updates, humor based on lesbian experience).
This localization reflects the agency of the native queer community in appropriating external cultural elements to suit their expressive needs and using habits \cite{ohiagu2014social}. 
On the other hand, \#wlw becomes a broader signifier encompassing feminism and general womanhood.
However, this expanded usage of \#wlw beyond queer identity risks diluting its specific function as a marker for sexual minorities. While such semantic diffusion may enhance visibility for women's rights, it can also obscure queer women and internal diversity within the them. Specifically, the identities of bi/pan-sexual, and transgender individuals—who are already marginalized within lesbian discourse \cite{marzetti2022really, nelson2023social} —may become further invisibilized under \#wlw's generalized use. This raises critical concerns about intersectional erasure in online queer culture, highlighting \textbf{the need for platforms and communities as human cyberinfrastructure to consciously maintain representational inclusivity when negotiating shared labels}.

By examining how two queer communities interpreted and interacted with the hashtag \#wlw during this event, our study offers valuable insights into cross-cultural communication and the identity formation of queer communities in online environments. \textbf{This case demonstrates how digital platforms can serve as arenas for intercultural negotiation, identity re-articulation, and community adaptation}. For future research, we plan to expand the sample size of our analysis by continuing temporal collection, further examining the comments, and conducting user interviews to incorporate user perspectives, allowing for a more comprehensive understanding of how users engage with and reinterpret the external hashtag \#wlw. Furthermore, while we examined the impact of \#wlw on RedNote, it would be valuable to explore whether TikTok users brought any hashtags or usage habits from RedNote back to TikTok. Such investigation could reveal the bidirectional nature of platform migration and how temporary cultural exchanges may leave lasting effects on both platforms' queer communities.

\bibliographystyle{ACM-Reference-Format}
\bibliography{sample-base}

\appendix
\onecolumn
\section{Appendix}
\begin{table}[htbp]
\caption{Codebook of the Post Content}
\label{tab:coding}
\resizebox{\columnwidth}{!}{%
\begin{tabular}{llll}
\hline
\textbf{Theme} &
  \textbf{Code} &
  \textbf{Definition} &
  \textbf{Sub-code} \\ \hline
\textit{\textbf{Commercial Content}} &
  \textit{Female Oriented} &
   &
   \\
 &
  \textit{General Audience Oriented} &
   &
   \\
\textit{\textbf{}} &
  \textit{Lesbian Oriented} &
  \multirow{-3}{*}{Commercial content advertising a merchandise or a profitable organization} &
   \\ \hline
\textit{\textbf{Influx Related Dialogues}} &
  \textit{Native-Initiated Dialogues with Immigrants} &
   &
  questions, comments \\
\textit{\textbf{}} &
  \textit{Immigrant-Initiated Dialogues with Natives} &
  \multirow{-2}{*}{Dialogues initiated by RedNote natives or TikTok immigrants regarding the influx} &
  questions, comments \\ \hline
\textit{\textbf{Knowledgeable Content}} &
  \textit{Critical Discussion} &
  Serious discussion about \#wlw and queer community related topics &
  \begin{tabular}[c]{@{}l@{}}\#wlw, coming out and identity, Chinese \\ queer community, queer concepts\end{tabular} \\
\textit{\textbf{}} &
  \textit{Feminism Content} &
  Content promoting feminism-related topics &
   \\
 &
  \textit{Lesbian Encyclopedia} &
  Everyday and serious knowledge about being a lesbian &
  \begin{tabular}[c]{@{}l@{}}instructions, queer knowledge introduction, \\ academic/fictional literature\end{tabular} \\ \hline
\textit{\textbf{Personal Showcase Content}} &
  \textit{Personal Lesbian Experience} &
  Personal experience of being a lesbian &
  \begin{tabular}[c]{@{}l@{}}love story, coming out and identity, \\ biphobia, living together, breakup, \\ sex, male gaze, stereotype, straight crisis\end{tabular} \\
\textit{\textbf{}} &
  \textit{Public Display of Affection} &
  Showing off the affection of lesbian couples &
   \\
\textit{\textbf{}} &
  \textit{Self-Promotion within Lesbians} &
  Promote themselves targeting lesbian audiences &
  \begin{tabular}[c]{@{}l@{}}daily life, appearance, thirst trap, talents \\ and skills, high-level reflection\end{tabular} \\ \hline
\textit{\textbf{Recreational Content}} &
  \textit{Lesbian Drawing} &
  {\color[HTML]{000000} Lesbian related painting creation} &
   \\
\textit{} &
  \textit{Lesbian Fandom Recreation} &
  Lesbian related fandom recreation (movie, TV shows, celebrities) &
   \\
\textit{} &
  \textit{Lesbian Humor} &
  Humors related to lesbian life and characteristic &
  \begin{tabular}[c]{@{}l@{}}Chinese wordplay, coming out, sex, \\ stereotype, male gaze, brainrot, illegalness \\ in some regions, relationship declaration\end{tabular} \\
 &
  \textit{Lesbian Story-writing} &
  Lesbian related story/fiction creation &
   \\
\textit{} &
  \textit{TikTok-style (Couple) Challenge} &
  TikTok-style challenges, including popular (lesbian) couple challenges &
   \\ \hline
\textit{\textbf{Socializing Content}} &
  \textit{Dating Request} &
  Requesting dates/relationship with (potential) girlfriend &
   \\
 &
  \textit{1 to n Networking} &
  Making networking/socializing requests to wider audience (not dating) &
  \begin{tabular}[c]{@{}l@{}}friend making, asking for recommendation, \\ organizing queer communication\end{tabular} \\
 &
  \textit{n to n Greetings from Queer Organizations} &
  Greetings from queer organizations to recruit new members &
   \\ \hline
\end{tabular}%
}
\end{table}

\end{document}